\documentclass[prl,twocolumn,amsmath,amssymb]{revtex4}

\usepackage{graphicx}
\begin{document}

\title{
Tunneling mediated by conical waves in a 1D lattice
}
\author{Andrea Di Falco,$^{1}$ Claudio Conti,$^{2,3}$ Stefano Trillo$^{3,4}$}
\email{claudio.conti@phys.uniroma1.it}
\affiliation{
$^{1}$ School of Physics and Astronomy, University of St. Andrews,
North Haugh, St. Andrews, KY16 9SS, UK\\
$^{2}$Centro Studi e Ricerche ``Enrico Fermi'', Via Panisperna 89/A, 
00184 Rome, Italy \\
$^{3}$Research center SOFT INFM-CNR Universit\`{a} di Roma ``Sapienza'',  P. A. Moro 
2, 00185, Roma, Italy\\
$^{4}$ Dipartimento di Ingegneria, Universit\`{a} di Ferrara, Via Saragat 1, 
44100 Ferrara, Italy
} 
\date{\today}
\begin{abstract}
The nonlinear propagation of 3D wave-packets in a 1D Bragg-induced band-gap system, 
shows that tranverse effects (free space diffraction) affect the interplay of periodicity and nonlinearity,
leading to the spontaneous formation of fast and slow conical localized waves. 
Such excitation corresponds to enhanced nonlinear transmission (tunneling) in the gap, 
with peculiar features which differ on the two edges of the band-gap, 
as dictated by the full dispersion relationship of the localized waves.
 \end{abstract}
\maketitle
Localized wave-packets (LW) characterized by a conical envelope, and in particular X waves, have been invoked as
a new paradigm that allows to explain several {\em nonlinear} optical wave phenomena
observed in dispersive media, such as fsec trapping via frequency doubling 
and filamentation in Kerr-like media \cite{DiTrapani03prl,Conti03prl,Kolesik04prl,Faccio06prl}, 
as well as spatio-temporal localization in lattices \cite{Lahini07prl}.
Unlike solitons, these LW exist also in the {\em linear}  limit where they are expressed
as superpositions of Bessel beams, well known in different context of physics \cite{Lu92,Porras04,book}.  
The nonlinearity "dresses" these solutions, 
playing a key role  (often strongly dynamical \cite{Conti03prl,Kolesik04prl}) 
in their formation and morphing process.

It has been suggested that LW can also exist as envelope modulations
of linear Bloch modes in periodic media such as 1D lattices (with free diffraction in the transverse direction) \cite{Conti04prl},
or 3D generalizations, namely photonic crystals \cite{Longhi04prb,Staliunas06}. 
In this case they play a major role around the forbidden band-gaps where, 
however, strong reflection is expected to severely hamper their formation in the linear regime.
While in principle the nonlinearity can be envisaged 
to favour the energy coupling and thus the
excitation of LW, the dynamics of this process is completely unexplored so far. 
Furthermore, one can argue that conical LW can compete with gap soliton bullets (GB)
\cite{GSbullets,Dohnal05,Sukh06prl}, which are the multi-dimensional analog 
(usually under more restrictive conditions) 
of well-known gap solitons of 1D settings  \cite{Chen87prl}, observed, 
e.g., as slow light pulses in fiber Bragg gratings \cite{Eggl96prl,Mok06nat},
spatial envelopes in array of coupled waveguides \cite{Mandelik04prl}, 
or atom clouds in a ultracold Bose condensed gas with standing wave optical lattice \cite{Eiermann04prl}.

In this Letter we investigate the dynamic role of LW owing to
additional free transverse dynamics (diffraction) included in a system
with stop-band (whose 1D longitudinal dynamics and gap solitons are well known \cite{Chen87prl,Eggl96prl,Mok06nat,Mandelik04prl,Eiermann04prl}). 
Employing excitations easily implementable in experiments 
(unidirectional and bell-shaped beams),
we show numerically that, in lattices of finite length, highly transmissive nonlinear behavior in the stop-band
is mediated by the excitation of conical LW. 
The features of this process can be understood on the basis of
the  linear multi-dimensional dispersion relationship of fast and slow LW, 
which entails a different dynamics on the two edges of the stopband.
While on the lower edge, LW have a primary role to determine the enhanced nonlinear transmission,
on the upper edge LW compete with GB, and their excitation stems from components
tunneling through the gap in transverse wave-number.
Thus our results suggest that gap solitons are not the sole entities that 
rule multi-dimensional nonlinear transmission through a stop-band.

We start from the following continuous model that rules (in dimensionless units) the propagation
around a single Bragg resonance of a period 1D system \cite{Chen87prl,Conti04prl,Dohnal05}
\begin{eqnarray}	
\label{CMT1}	
		\left(i \partial_{t} + i \partial_{z} + \frac{1}{2} \nabla_\perp^2 \right) u_{+} + u_{-} +
		\chi \left(\left|u_{+}\right|^2+2\left|u_{-}\right|^2\right)u_{+}=0, \nonumber \\
		 \\
		\left(i \partial_{t} - i \partial_{z} + \frac{1}{2} \nabla_\perp^2 \right) u_{-} + u_{+} +
		 \chi \left(\left|u_{-}\right|^2+2\left|u_{+}\right|^2\right)u_{-}=0,\nonumber 
\end{eqnarray}
\noindent where $t$ and $z$ are normalized time and distance (along the 1D lattice),  
and $\nabla_\perp^2=\partial_x^2+\partial_y^2$ is the transverse Lapacian.
Though we expect a similar scenario with radially symmetric
beams in two transverse dimensions, we discuss the computationally less demanding case of one transverse dimension
($\nabla_\perp^2=\partial_y^2$).  
In this case Eqs. (\ref{CMT1}) hold for Bragg optical gratins in a slab waveguide,
or for Bose-Einstein atom condesates with free kinetic spreading (no trap in $y$)
transversally to a standing wave potential $\sin^2 (k_{g} z/2)$.
In the former case $t,z,y$	 are measured in units of $T_{0}=(V_{g} \Gamma)^{-1}$,
$Z_{0}=\Gamma^{-1}$, $Y_{0}=(2k_{0} n_{0} \Gamma)^{-1/2}$, respectively, whereas
$u_{\pm}= \sqrt{k_{0} |n_{2I}|/2\eta} E_{\pm}$,  $E_{\pm}$ are counterpropagating envelopes at Bragg frequency, 
$k_{0}$ and $\eta$ are the relative vacuum wavenumber and impedance
of the host medium, $\Gamma$ is the Bragg coupling coefficient,
and $n_{2I}$ the Kerr nonlinear index. In the latter case, 
Eqs. (\ref{CMT1}) derive from the mean-field Gross-Pitaevskii (GP) equation \cite{Conti04prl} with
$T_{0}= \hbar/\Gamma$, $Z_{0}^2= \hbar^{2} k_g/(2 m\Gamma)$,
$Y_{0}^2=\hbar^{2}/m \Gamma$, $m$ being the atomic mass
whereas $4\Gamma$ is the peak energy of the periodic potential.
$\psi_{\pm}=a^{-1/2} u_{\pm}$ are mutually Bragg scattered components of the atomic wave-function, 
with $a=(4\pi \hbar^{2}/m)a_{s}$, $a_{s}$ being the scattering length.
In the following we restrict the discussion to the focusing (or attractive) case
i.e. $\chi=sign(n_{2I})=-sign(a_{s})=1$, though similar phenomena can occur also in the defocusing
(repulsive) case.

The reconstruction of the overall field or wave function 
(apart from inessential common phase shifts) 
$u=u_{+} \exp \left(i k_g z/2 \right) + u_{-} \exp \left(-i k_g z/2 \right)$,
shows the nature of the envelopes $u_{\pm}$ as modulations of Bloch modes.
\begin{figure}
\includegraphics[width=8.5cm]{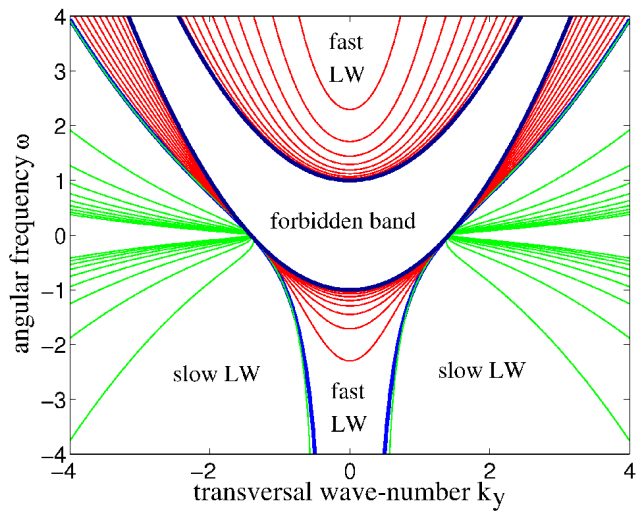}
\caption{
\label{figLWspectrum}
(Color online) 
Spatio-temporal dispersion curves $k_z=\omega/v$ of LW in the frequency plane $k_y,\omega$.
Outside the forbidden band (region between the two parabolas),
the LW light-line ($v=1$, blue curve) divides the regions of
subluminal LW ($v<1$, green thin curves) and superluminal LW ($v>1$, thick red curves).
}
\end{figure}
\paragraph{Dispersion and localized wave spectra. ---}
The linear dispersion [$u_{\pm} \sim \exp(ik_z z+i k_y y-i \omega t)$]
relationship associated with Eqs. (\ref{CMT1}) reads as
$k_z^2=(\omega-k_y^2/2)^2-1$, which entails a lower (LB) and upper branch (UB)
separated by a forbidden band in $\omega$  (or energy). 
This gap of edges $\omega_{\pm}=k_y^2/2 \pm 1$,
has constant width and central frequency shifting upward (parabolically) with $k_y$
(tilted waves with $k_y \neq 0$ see an effectively shorter period).
While the UB $\omega_+(k_y,k_z)$ has always positive curvature,
it is near the saddle-shaped (i.e., $d^2 \omega/dk_z^2 d^2 \omega/dk_y^2 < 0$)
LB $\omega_-(k_y,k_z)$ that X-shaped LW
has been predicted on the basis of a multiscale perturbative approach \cite{Conti04prl}.
However, at the linear level ($\chi=0$), more general existence conditions
for non-dispersive LW require simply a linear dependence $k_z=\omega/v$, 
where the arbitrary parameter $v=dk_z/d\omega^{-1}$ stands for
the group velocity of the LW travelling undistorted \cite{Conti03prl}. 
Upon substitution in the dispersion relationship we obtain
the dispersion curves of LW, namely $(\frac{\omega}{v})^2=(\omega-k_y^2/2)^2$,
shown in the plane $(k_y,\omega)$ (see Fig. \ref{figLWspectrum}) 
for various values of $v$ below and above
the luminal curve $v=1$, denoted henceforth as the {\it LW light-line}. 
The dispersion curves do not involve
evanescent waves and lie outside the gap (white region bordered by parabolas).
A wave-packet with spectrum along one of such curves represents a LW with well defined velocity
\footnote{The LW $(\psi_+,\psi_-)$ can be expressed as integral superposition
of Bessel functions, as we will show elsewhere. There is no explicit expression for such integrals.},
while quasi-stationary superpositions of LW \cite{Conti04prl} with defined mean velocity
result from spectra well concentrated in the different regions of Fig. \ref{figLWspectrum}.
In the nonlinear regime, as we show below, the nonlinearity acts as the triggering mechanism
of the spectral reshaping of the input into such type of LW.
We point out that, though LW exist above the UB edge where they correspond to
$O-waves$ \cite{Porras04}, for excitation in the band-gap (regardless whether in proximity of the LB or UB)
those playing a role are those below the LB edge, which are of the X type \cite{Porras04}
due to the opposite curvatures.
\paragraph{Numerical method. ---}
We consider the boundary value problem given by Eqs. (\ref{CMT1}) subject to 
unidirectional (forward) gaussian excitation
\begin{equation} \label{bc}
u_{+}(t,x,z=-L)=u_0 \exp \left[-\frac{(t-t_s)^2}{t_0^2} -\frac{y^2}{y_0^2} \right] \exp(-i \bar{\omega} t)
\end{equation}
and $u_{-}(t,y,z=L)=0$,
where $ \bar{\omega}$ is the frequency (energy) detuning that fixes the position in the gap,
and $E=\pi \,u_0^2\, t_0^2\, y_0^2$ is the input square norm (fluence).
Though we have performed extensive numerical runs, we show below
a typical case obtained for $2L=4$, $t_s=30$, $t_0=12$, $y_0=6$.
Since Eqs. (\ref{CMT1}) and (\ref{bc}) cannot be solved by standard
beam-propagation tecniques \cite{agrawal_book} (the main difficulty arises from 
the presence of a finite integration region in the propagation direction $z$ and an
infinite transverse domain in $y$), we adopted a pseudo-spectral technique
based on a Chebishev-collocation approach along $z$,
in conjuction with second-order split-step  (evolution variable $t$,
Fast-Fourier trasform in $y$) \cite{spectral_methods}.

\begin{figure}
\includegraphics[width=8.5cm]{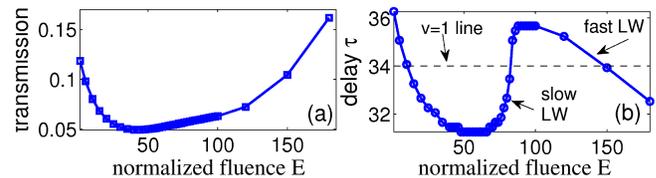}
\caption{
\label{figTLB}
(Color online) Excitation close to LB ($\bar \omega=-0.99$).
(a) Transmission Vs input fluence $E$; 
(b) Output pulse delay $\tau$ Vs $E$. 
The delay is calculated as the position
of the first peak in the transmitted $y-$integrated pulse profile.
}
\end{figure}

\begin{figure}
\includegraphics[width=8.3cm]{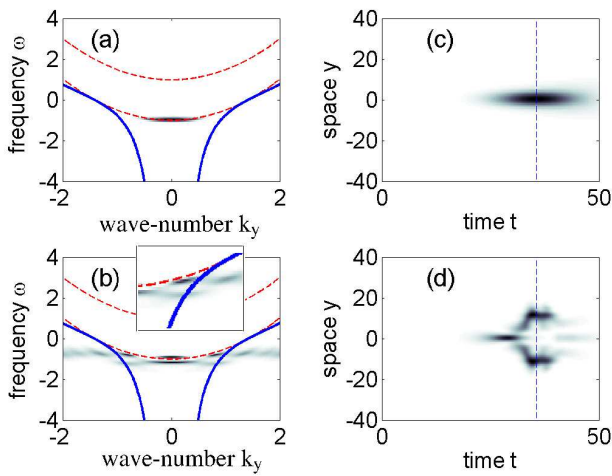}
\caption{
\label{figLBlong}
(Color online) 
(a,b) Output spectrum of $u_+$ at (a) low fluence $E=1$,
(b) high fluence $E=120$ (inset: zoom around $k_y\cong 1$). 
The red dashed line denotes the forbidden band, 
the blue thick line is the LW light-line.
(c,d) corresponding output spatio-temporal profile 
for (c) $E=1$; (d) $E=120$. 
The vertical dashed line gives the reference instant of the peak of $|u_+|^2$ at $E=1$.
}
\end{figure}

\begin{figure}
\includegraphics[width=8.3cm]{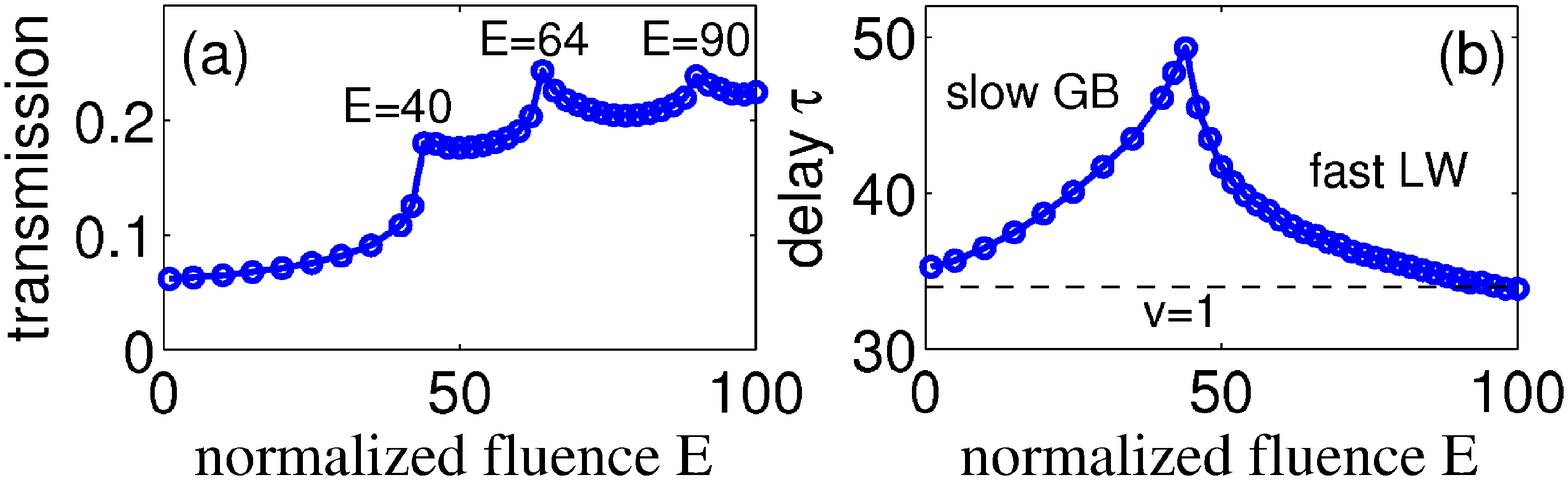}
\caption{
(Color online) As in Fig. \ref{figTLB} for $\bar \omega=0.99$ (LB).
}
\label{figTUB}
\end{figure}

\paragraph{Lower branch excitation. ---}
In Fig. \ref{figTLB}a we show the transmission $\mathcal{T}$
for an excitation inside the gap in proximity of the LB. 
At low fluence $E$ there is a residual transmission (owing to the finite size of the system),
which decreases at moderate fluences consistently with the nonlinear shift of the gap toward
lower frequencies (energies). However, at sufficiently high fluences, 
the trasmission increases with $E$. This nonlinear self-transparency
cannot be ascribed to stable GB, which do not exist close to the LB  \cite{Dohnal05} 
(on the LB gap soliton are unstable even in the plane wave limit $\nabla_y^2=0$).
However, the phenomenon can be explained by looking at the spectra
and $t-y$ snapshots (top view) of the fields at the output ($z=L$).
In Fig. \ref{figLBlong} we compare the case of low ($E=1$) and high ($E=120$) fluence.
While the output $k_y-\omega$ spectrum at $E=1$ is well concentrated
in the gap around small wave-numbers (Fig. \ref{figLBlong}a), 
at high fluence the strong self-focusing causes the spatial spectrum to widen
and fill the region of large wave-numbers (Fig. \ref{figLBlong}b).
As also evident in the inset, the energy is progressively concentrated in proximity 
of the LW light-line, with a clear peak in the region of fast LW, arising
from tunneling in wave-number outside the band-gap.
Correspondingly the $t-y$ snapshots show an almost undistorted (nearly gaussian)
beam at low fluences (Fig. \ref{figLBlong}c), which at large fluences, develops 
clear conical tails along the trailing edge (Fig. \ref{figLBlong}d).
Importantly the leading edge of the pulse arrives earlier in the nonlinear case
consistently with the spectral peak in the region of fast LW.
However, also spectral components in the region of slow LW are present
as clear from the inset in Fig. \ref{figLBlong}b. The interplay between these two components explains
the behavior of the output delay vs. $E$ displayed in Fig. \ref{figTLB}b. 
As $E>50$, the enhanced transparency of the structure is accompained 
by a growing (with $E$) delay. This corresponds to the excitation of slow LW components (see Fig. \ref{figLBlong}b).
However we also found that energy is coupled to frequencies
in the region between the lower band-edge and the LW
light-line (see inset in Fig. \ref{figLBlong}b).
The latter process becomes dominant at large $E$, being responsible
for the delay to decrease with $E$, for $E>100$. 
In the spatio-temporal domain, the fast LW corresponds
to a bell-shaped 
leading edge of the transmitted pulse, 
while the slow LW can be associated to slow conical tails, 
as shown in Fig. \ref{figLBlong}d.
\paragraph{Upper branch excitation. ---}
The excitation in the proximity of the UB is characterized by the existence of GB
which provide the main tunnelling mechanism \cite{Dohnal05}.
Evidence of this process is reported in Fig. \ref{figTUB}a, showing that
the trasmission $\mathcal{T}$ increases abruptly around $E\simeq 40$.
Above this threshold, $\mathcal{T}$ remains high, while it exhibits clear peaks. 
In addition, as shown in Fig. \ref{figTUB}b, the delay versus fluence $E$ 
displays a cusp in correspondence with the threshold. 
This implies that the generated nonlinear waves have a decreasing 
velocity for increasing fluence up to the threshold value, 
after which this trend is reversed.

\begin{figure}
\includegraphics[width=8.3cm]{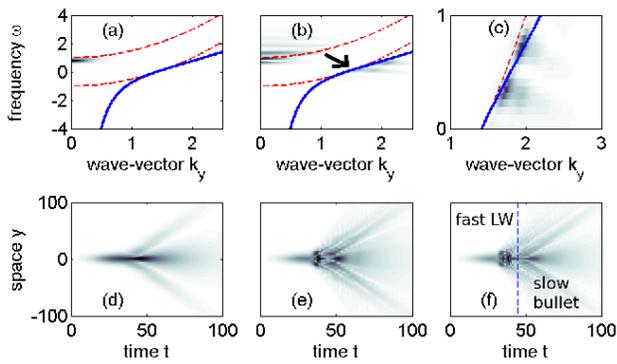}
\caption{(Color online) Output features of $u_+$
for UB excitation ($\bar \omega=0.99$).
Output spectra (a) $E=40$, (b) $E=64$, (c) $E=90$ (zoom around LW light-line); 
Corresponding intensity $(t,y)$ profiles: (d) $E=40$, (e) $E=64$, (f) $E=90$.
The dashed line in (e,f,g) gives the reference instant corresponding
to peak of $|u_+|^2$ at $E=40$.
\label{figUBlong}} \end{figure}

\begin{figure}
\includegraphics[width=8.3cm]{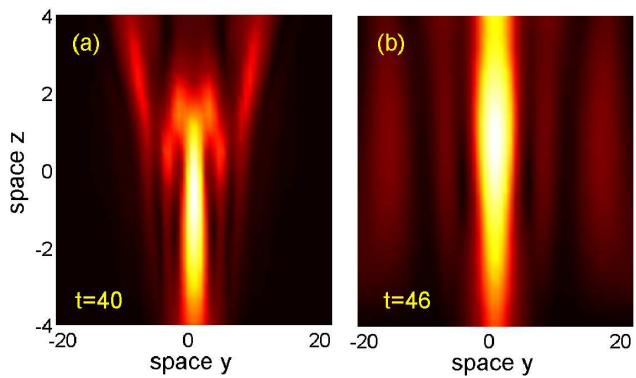}
\caption{(Color online) 
Level plot of spatial intensity $|u_+|^2+|u_-|^2$
for UB excitation ($\bar \omega=0.99$), at two instants: 
(a) $t=40$, both the fast conical LW and the GB are trapped
in the lattice; (b) $t=46$, the fast LW has left the lattice 
and only the slow GB remains.
\label{figZY}} \end{figure}

This dynamics is a direct consequence of the multi-dimensional nature of the problem and
the existence of LW solutions. In Fig. \ref{figUBlong}a,d we show the
$k_y-\omega$ spectrum and $(y,t)$ snapshot at the output $z=L$, 
in correspondence of the first peak in the transmission $\mathcal{T}$.
As shown, the tunneling wave-packet has a spectrum which remains well confined inside the gap
around low wave-numbers. Correspondingly the beam, which has undergone self-focusing
in the transverse variable, has substantially bell-shaped section (elliptical top view)
with very weak V-shaped tails.
Such evolution indicates that the transmission peak is basically associated with the
formation of a GB tunneling through the lattice.
The situation remains nearly unchanged until the second peak of transmission at $E=64$ is reached.
Here, the situation is radically different as clear from the spectrum
in Fig. \ref{figUBlong}b. Owing to strong self-focusing, a component of the spectrum
has clearly tunneled (mainly horizontally, i.e. in $k_y$) through the gap to the LB.
In $t,y$ domain this spectral feature has one to one correspondence
(as can be shown by filtering and anti-transformation) with the strong enhancement
of the V-shaped tails characteristic of X waves (see Fig. \ref{figUBlong}e).
Since the LB spectral component lies in the region of fast LW (i.e., the tiny
region between the LW light-line and the band-edge), we conclude that the transmission
peak is associated with the concomitant excitation of a such type of waves. 
The fast character of the LW of the X type is also clear from Fig. \ref{figUBlong}e, which
shows that the tails emanate from a wave component that anticipates the slow GB component.
The tunnelling process due to the generation of the fast X-wave  goes on
until the fluence is increased up to the next transmission peak
at $E=90$ (see Fig. \ref{figTUB}a), where a second LW is generated,
as shown by the splitting of the LB spectral component (see Fig. \ref{figUBlong}c),
and the more structured fast wave component in Fig. \ref{figUBlong}f.
The excitation of a fast LW explains the cusp in the delay time $\tau$ in Fig. \ref{figTUB}b:
as energy is increased the bullet develops fast LW tails that eventually precede the
GB and correspondingly (cusp in Fig. \ref{figTUB}b) the delay starts to decrease with $E$.
This process is even more evident in Fig. \ref{figZY}, where we show the intensity profile for $E=50$
and $2L=8$ (a longer structure makes the process more evident), 
at two different instants. For $t=40$, the intensity distribution 
clearly unveils a X-shaped pulse which is in proximity of the output.
At $t=46$ the fast LW has left the structure, and only the slow GB remains.
\paragraph{Conclusions. ---}
In summary the study of transverse effects a 1D Bragg 
lattice unveils the role of fast and slow LW,
characterized by specific spatio-temporal spectra.
Their excitation foster a variety of self-trasparency effects, 
either in conjuction with multi-dimensional gap solitons,
or as an alternative to them. 
Measuring the spatio-temporal spectra or the trend of the output delay with 
input fluence could be used to monitor the excitation of LW.
Our results have implication in all of the field where multi-dimensional
solitary wave and periodicity play a role, and more specifically in Bose Einstein 
condensation and nonlinear optics.

\begin{acknowledgments}
We thank INFM-CINECA initiative for parallel computing
and MIUR for financial support. A.D.F. is supported by the Consortium of Speckled Computing (EPSRC ``Specknet'').
\end{acknowledgments}


\begin{thebibliography}{30}


\bibitem{DiTrapani03prl} 
P. Di Trapani {\it et al.}, 
Phys. Rev. Lett. {\bf 91}, 093904 (2003).

\bibitem{Conti03prl}
C. Conti {\it et al.}, 
Phys. Rev. Lett. {\bf 90}, 170406 (2003);
C. Conti,
Phys. Rev. E {\bf 70}, 046613 (2004).

\bibitem{Kolesik04prl} 
M. Kolesik, E.M. Wright, and J. V. Moloney, 
Phys. Rev. Lett.  {\bf 92}, 253901 (2004).

\bibitem{Faccio06prl} 
D. Faccio, M. Porras, A. Dubietis, F. Bragheri, A. Couairon, P. Di Trapani, 
Phys. Rev. Lett. {\bf 96}, 193901 (2006).

\bibitem{Lahini07prl} 
Y. Lahini, E. Frumker, Y. Silberberg, S. Droulias, K. Hizanidis, and D. N. Christodoulides,
Phys. Rev. Lett. {\bf 98},  023901 (2007);
S. Droulias, K. Hizanidis, J. Meier, and D. N. Christodoulides,
Opt. Exp. {\bf 13}, 1827 (2005).

\bibitem{Lu92}
J.~Lu and J.~F. Greenleaf,
IEEE Trans. Ultrason. Ferrelec. Freq. contr. \textbf{39}, 441 (1992);
P.~Saari and K.~Reivelt,
\prl \textbf{79}, 4135 (1997).

\bibitem{Porras04} M. Porras and P. Di Trapani
Phys. Rev. E {\bf 69}, 066606 (2004).

\bibitem{book}
E. Recami, M. Zamboni-Rached, and H.E. Hernandez-Figueroa,
{\em Localized waves} (Wiley, 2007).

\bibitem{Conti04prl}
C. Conti, S. Trillo,
Phys. Rev. Lett. {\bf 92}, 120404 (2004).


\bibitem{Longhi04prb} 
S. Longhi and D. Janner, 
Phys. Rev. B {\bf 70}, 235123 (2004);
S. Longhi, 
Phys. Rev. E {\bf 71}, 016603 (2005).

\bibitem{Staliunas06} 
K. Staliunas and R. Herrero, 
Phys. Rev. E {\bf 73}, 016601 (2006).

\bibitem{GSbullets} 
S. John and N. Ak\"{o}zbek,
Phys. Rev. Lett. {\bf 71}, 1168 (1993);
A. Aceves, B. Costantini, C. De Angelis,
J. Opt. Soc. Am. B {\bf 12}, 1475 (1995);
N. Ak\"{o}zbek, and S. John,
Phys. Rev. E {\bf 57}, 2287 (1998);
A. B. Aceves, G. Fibich, B. Ilan,
Physica D {\bf 189}, 277 (2004).

\bibitem{Dohnal05} 
T. Dohnal, A. B. Aceves, 
Stud. Appl. Math. {\bf 115}, 209 (2005).


\bibitem{Sukh06prl} A. Sukhorukov and Y. S. Kivshar,
Phys. Rev. Lett. {\bf 97},  233901 (2006).

\bibitem{Chen87prl} 
W. Chen and D. L. Mills, Phys. Rev. Lett. {\bf 58}, 160 (1987);
C. M. de Sterke and J. E. Sipe, Phys. Rev. A {\bf 39}, 5163 (1988);
A. B. Aceves and S. Wabnitz, Phys. Lett. A {\bf 141}, 37 (1989);
D. N. Christodoulides and R. I. Joseph, Phys. Rev. Lett. {\bf 62}, 1746 (1989).

\bibitem{Eggl96prl}  
B. J. Eggleton, R. E. Slusher, C. M. de Sterke, P. A. Krug, and J. E. Sipe,
Phys. Rev. Lett. {\bf 76}, 1627 (1996).

\bibitem{Mok06nat} 
J. T. Mok,  C. M. De Sterke, I. C. M. Littler, and B. J. Eggleton,
Nature Phys. {\bf 2}, 775 (2006).

\bibitem{Mandelik04prl} 
D. Mandelik, R. Morandotti, J.S. Aitchison, and Y. Silberberg, 
Phys. Rev. Lett. {\bf 92},  093904 (2004).
D. Neshev, A. Sukhorukov, B. Hanna, W. Krowlikowsky, and Y. S. Kivshar,
Phys. Rev. Lett. {\bf 93},  083905 (2004).

\bibitem{Eiermann04prl}
S.~P\"{o}tting, P.~Meystre, and E.~M.Wright, in
\emph{Nonlinear photonic crystals}
(Springer, 2003), vol.~10 of \emph{Photonics};
B. Eiermann {\it et al.}, 
Phys. Rev. Lett. {\bf 92}, 230401 (2004).
O. Morsch and M. K. Oberthaler, Rev. Mod. Phys.  {\bf 78}, 179 (2006).

\bibitem{agrawal_book}
G. P. Agrawal,
{\em Nonlinear fiber optics} 3rd et. (Academic Press, San Diego, 2001)

\bibitem{spectral_methods}  
J. P. Boyd, 
{\em Chebyshev and Fourier Spectral Methods} II ed. (Dover, New York, 2001);
C. Canuto, M. Y. Hussaini, A. Quaternoni and T. A. Zhang,
{\em Spectral Methods for Fluid Dynamics} (Springer-Verlag, New York, 1988)

\end{thebibliography}
\end{document}